\documentclass[prd,twocolumn,showpacs,showkeys,footinbib]{revtex4}
\usepackage{amsmath}
\usepackage{graphics}

\renewcommand\vec[1]{\mathbf{#1}}
\newcommand\he[1]{#1^{\dagger}}

\DeclareMathOperator{\tr}{tr}
\DeclareMathOperator{\Tr}{Tr}
\newcommand\D{\mathcal D}
\newcommand\bra[1]{\langle#1|}
\newcommand\ket[1]{|#1\rangle}
\begin{document}

\title{Spontaneous symmetry breaking in the linear sigma model\\
at finite chemical potential: One-loop corrections}
\author{Tom\'a\v s~Brauner}
\email{brauner@ujf.cas.cz}
\affiliation{Department of Theoretical Physics,
Nuclear Physics Institute, 25068 \v Re\v z, Czech Republic}

\begin{abstract}
We investigate spontaneous symmetry breaking within the linear sigma model with
the $\mathrm{SU(2)}\times\mathrm{U(1)}$ internal symmetry at finite chemical
potential, which was suggested as a~model for kaon condensation in the CFL
phase of dense quark matter. One-loop corrections to the scalar field effective
potential as well as its propagator are calculated. Particular attention is
paid to the type-II Goldstone boson that appears in the Bose--Einstein
condensed phase. Furthermore, we show that the type-I Goldstone boson --- the
superfluid phonon --- is allowed to decay due to the nonlinearity of its
dispersion relation at high momentum, and determine its decay width.
\end{abstract}

\maketitle

\section{Introduction}
One of the most striking and general consequences of spontaneous symmetry
breaking is the existence of gapless modes, guaranteed by the celebrated
Goldstone theorem \cite{Goldstone:1961eq,Goldstone:1962es}. The number and
properties of the Goldstone bosons are crucial for the low-energy dynamics (and
the low-temperature thermodynamics) of the system. In particular, these soft
modes play a~major role in transport phenomena such as heat conductivity or
viscosity.

It has been known for a~long time that the physics of spontaneous symmetry
breaking in Lorentz-noninvariant systems may be quite intricate. The
fundamental result in this respect was achieved by Nielsen and Chadha
\cite{Nielsen:1976hm}: They showed in a nonperturbative manner that, under
certain technical assumptions, the energy of the Goldstone bosons stemming from
the spontaneous symmetry breaking is, in the long-wavelength limit,
proportional to some power of momentum. The Goldstone bosons are then
classified as type-I, if this power is odd, and type-II, if it is even,
respectively. The numbers of the Goldstone bosons of the respective types are
related to the number of broken symmetry generators by the following inequality,
\begin{multline}
\text{\# type-I GBs}+2\times\text{\# type-II GBs}\\
\geq\text{\# broken generators}.
\label{Nielsen_Chadha}
\end{multline}

Eq. \eqref{Nielsen_Chadha} shows that whenever there is a type-II Goldstone
boson, the number of Goldstone bosons may be smaller than the number of broken
generators. A~profound example of a~system where this happens is provided by
the ferromagnet. In the past decade, however, several other systems with
type-II Goldstone bosons have been studied, ranging from various high-density
phases of QCD
\cite{Miransky:2001tw,Schaefer:2001bq,Buballa:2002wy,Blaschke:2004cs} and the
relativistic nuclear ferromagnet \cite{Beraudo:2004zr} to Bose--Einstein
condensed atomic gases \cite{Ho:1998ho,Ohmi:1998om,He:2006ne} and the general
effect of relativistic vector condensation
\cite{Sannino:2001fd,Sannino:2002wp}.

In our recent paper \cite{Brauner:2005di} we investigated the Goldstone boson
counting in a~particular class of Lorentz-noninvariant systems --- the
relativistic linear sigma model at finite chemical potential. Based on previous
partial results \cite{Leutwyler:1994gf,Schaefer:2001bq} we clarified the
connection of Goldstone boson counting with the possibility of nonzero
densities of Noether charges and conjectured a~general counting rule:
\emph{Non-zero density of a~commutator of two broken generators implies one
type-II Goldstone boson with a~quadratic dispersion law}. As a consequence, the
inequality \eqref{Nielsen_Chadha} is saturated, up to exceptional cases like
the phase transitions, where the phase velocity of a type-I Goldstone boson may
vanish, thus making it an `accidental' type-II one. Within the linear sigma
model, we were able to give a detailed proof of these statements.

Our analysis was, however, purely classical --- we worked all the time at the
tree level. The present paper is intended to fill this gap. There are several
reasons that make the inclusion of the loop corrections necessary. First, it is
only at the one-loop level that the $\lambda(\he\phi\phi)^2$ interaction plays
a~nontrivial role. Indeed, at the tree level, it has no effect on the spectrum
in the normal phase, while in the Bose--Einstein condensed phase it merely
serves to stabilize the scalar field potential. Second, the loop corrections
may play a~significant role, especially in the vicinity of the phase
transition, where the classical analysis cannot be trusted. Third, it is well
known that when the static part of the Lagrangian has higher symmetry than the
full Lagrangian, the tree level is not sufficient to determine the spectrum
(even qualitatively) correctly \cite{Coleman:1973jx}. The quantum corrections
are therefore necessary in order to check the saturation of the Nielsen--Chadha
inequality \cite{Nielsen:1976hm}.

For simplicity, we do not follow the general symmetry-breaking pattern of Ref.
\cite{Brauner:2005di}. Instead, we analyze a~particular model --- the linear
sigma model with the global $\mathrm{SU(2)}\times\mathrm{U(1)}$ symmetry. This
model has been used to describe kaon condensation in the Color-Flavor-Locked
phase of dense QCD \cite{Miransky:2001tw,Schaefer:2001bq}. It is also a~special
case of a~class of models of relativistic Bose--Einstein condensation
investigated by Andersen \cite{Andersen:2005yk}.

The plan of the paper is as follows. Using the manifestly invariant formalism
of generating functionals, we determine in Sec. \ref{Sec:effective_potential}
the one-loop effective potential and numerically find its minimum as a~function
of the chemical potential. In Sec. \ref{Sec:scalar_propagator}, we calculate
the one-loop correction to the scalar field propagator. We demonstrate
analytically the presence of gapless (Goldstone) poles and then compute
numerically the corrections to the tree-level dispersion relations of the
type-II Goldstone boson and its massive counterpart. In Sec.
\ref{Sec:phonon_decay}, we calculate the decay rate of the superfluid phonon
\cite{Son:2002zn}. The paper is concluded with a~summary and a~discussion of
the results. Some necessary technical details are deferred to the appendices.

\section{Effective potential}
\label{Sec:effective_potential} We work with the model of Miransky and Shovkovy
\cite{Miransky:2001tw} and Schaefer \emph{et al.} \cite{Schaefer:2001bq} which
is defined by the Lagrangian
\begin{equation}
\mathcal L=\he{D_{\mu}\phi}D^{\mu}\phi-M^2\he\phi\phi-\lambda(\he\phi\phi)^2.
\label{Lagrangian}
\end{equation}
Here $\phi$ is a~complex doublet of the global $\mathrm{SU(2)}$ symmetry and
the chemical potential $\mu$, associated with the global $\mathrm{U(1)}$
symmetry (particle number), is included in the covariant derivative,
$D_{\mu}\phi=(\partial_{\mu}-i\delta_{\mu0}\mu)\phi$ \cite{Kapusta:1989ka}.
This model has been analyzed, at the tree level, in detail in Refs.
\cite{Miransky:2001tw,Schaefer:2001bq}, so we only summarize the main results
for later reference.

When $\mu>M$ the static potential develops a~nontrivial minimum and the scalar
field condenses. As a~result, the $\mathrm{SU(2)\times U(1)}$ symmetry of the
Lagrangian \eqref{Lagrangian} breaks down to its $\mathrm{U(1)}$ subgroup
[different from the original $\mathrm{U(1)}$]. Associated with the \emph{three}
broken generators there are \emph{two} Goldstone bosons, one type-I and one
type-II. Their low-energy dispersion relations read
\begin{equation}
E=\sqrt{\frac{\mu^2-M^2}{3\mu^2-M^2}}|\vec p|\quad
\text{and}\quad
E=\frac{\vec p^2}{2\mu},
\label{disp_rel_tree_level}
\end{equation}
respectively. In addition to the Goldstone bosons, there are two massive
excitations with energy gaps $\sqrt{2(3\mu^2-M^2)}$ and $2\mu$.

\subsection{One-loop correction}
Let us now proceed to the one-loop calculation. In order to account for the
breaking of the $\mathrm{U(1)}$ symmetry associated with the particle number (or
strangeness, in the context of kaon condensation) we introduce the formal Nambu
doublet
$$
\Phi_i=\left(
\begin{array}{c}
\phi_i \\ \he\phi_i
\end{array}\right).
$$
[The index $i$ refers to the doublet representation of the $\mathrm{SU(2)}$.]
The one-loop effective action is then given by the textbook formula
\cite{Peskin:1995ev},
\begin{equation}
\Gamma_{\text{1L}}[\phi,\he\phi]=S[\phi,\he\phi]+\frac
i2\log\det\Delta^{-1}
\label{1loop_action}
\end{equation}
plus counterterms. Here $S[\phi,\he\phi]=\int d^4x\,\mathcal L(x)$ is the classical
action and $\Delta$ is the tree-level matrix propagator defined by
$$
\Delta^{-1}_{ij}(x,y)=\frac{\delta^2S}{\delta\he\Phi_i(x)\delta\Phi_j(y)}.
$$
Displaying explicitly the matrix structure in the Nambu space, it reads, in the
momentum representation,
\begin{widetext}
\begin{equation}
\Delta^{-1}_{ij}(p)=\left(
\begin{array}{cc}
\bigl[(p_0+\mu)^2-\vec
p^2-M^2-2\lambda\he\phi\phi\bigr]\delta_{ij}-2\lambda\he\phi_{j}\phi_{i} &
-2\lambda\phi_{i}\phi_{j}\\
-2\lambda\he\phi_{i}\he\phi_{j} &
\bigl[(p_0-\mu)^2-\vec
p^2-M^2-2\lambda\he\phi\phi\bigr]\delta_{ij}-2\lambda\he\phi_{i}\phi_{j}
\end{array}\right).
\label{propagator_tree_level}
\end{equation}
\end{widetext}
Here the classical field $\phi$ is already assumed to be constant. There is no
lack of generality in this requirement as long as the vacuum is translationally invariant.

With the constant classical field $\phi$ we may evaluate the one-loop effective
potential as
\begin{equation}
V_{\text{1L}}=V_{\text{cl}}-\frac
i2\int\frac{d^4k}{(2\pi)^4}\log\det\Delta^{-1}(k)
\label{1loop_potential}
\end{equation}
plus counterterms, where
$V_{\text{cl}}=(M^2-\mu^2)\he\phi\phi+\lambda(\he\phi\phi)^2$.

In this form, the effective potential is manifestly $\mathrm{SU(2)\times U(1)}$
invariant. For detailed calculations it is, however, more convenient to fix the
direction of $\phi$ in the $\mathrm{SU(2)}$ `flavor' space. As usual, we set
\begin{equation}
\phi=\frac1{\sqrt2}\left(
\begin{array}{c}
0 \\
v~\end{array}\right).
\label{vacuum_choice}
\end{equation}

With this choice, $\Delta^{-1}$ becomes diagonal in the flavor space and may easily
be inverted, yielding
\begin{widetext}
\begin{equation}
\begin{split}
\Delta_{11}(p)&=\left(
\begin{array}{cc}
\bigl[(p_0+\mu)^2-\vec p^2-M^2-\lambda v^2\bigr]^{-1} & 0\\
0 & \bigl[(p_0-\mu)^2-\vec p^2-M^2-\lambda v^2\bigr]^{-1}
\end{array}\right),\\
\Delta_{22}(p)&=\frac1{[p_0^2-E_+^2(\vec p)][p_0^2-E_-^2(\vec p)]}\left(
\begin{array}{cc}
(p_0-\mu)^2-\vec p^2-M^2-2\lambda v^2 & \lambda v^2\\
\lambda v^2 & (p_0+\mu)^2-\vec p^2-M^2-2\lambda v^2
\end{array}\right),
\end{split}
\label{Delta}
\end{equation}
\end{widetext}
where
\begin{multline*}
E^2_{\pm}(\vec p)=\vec p^2+\mu^2+M^2+2\lambda v^2\\
\pm\sqrt{4\mu^2(\vec p^2-\mu^2
+M^2+\lambda v^2)+(2\mu^2+\lambda v^2)^2}.
\end{multline*}
By minimizing the classical potential $V_{\text{cl}}$ we find that, at tree
level, $\lambda v^2=\mu^2-M^2$, which yields the dispersion relations
\begin{equation}
E^2=\vec p^2+3\mu^2-M^2\pm\sqrt{4\mu^2\
\vec p^2+(3\mu^2-M^2)^2},
\label{disp_rel_tree_level_exact}
\end{equation}
in accord with the result of Refs.
\cite{Miransky:2001tw,Schaefer:2001bq}.

\subsection{Renormalization}
The loop integral in Eq. \eqref{1loop_potential} is divergent. To renormalize
it, we add to the effective potential the counterterm
$$
V_{\text{counterterm}}=\delta v+\delta
M^2\he\phi\phi+\delta\lambda(\he\phi\phi)^2.
$$
This term generates, by means of its functional derivatives, also the
counterterm for the scalar propagator. The constant $\delta v$ is irrelevant
for our purposes as it just fixes the offset of the effective potential at
$\phi=0$. The counterterms in the effective potential are sufficient to absorb
all infinities for, as is well known, there is no wave function renormalization
in the scalar $\phi^4$ theory at one loop \footnote{Strictly speaking, this is
only true for the normal phase. In the Bose--Einstein condensed phase, there is
a~\emph{finite} wave function renormalization induced by the second term in the
integral in Eq. \eqref{full_propagator}. This wave function renormalization is,
in fact, responsible for the renormalization of the Goldstone boson dispersion
relations.}.

Actually, we shall not need to evaluate the effective potential at all. What we
are interested in is just the vacuum expectation value of the scalar field, or
$v$. We shall be therefore solving the `gap equation' $\partial
V_{\text{1L}}/\partial v=0$. Now since the effective potential is a~function of
$v^2$, there is always the solution $v=0$. In the Bose--Einstein condensed
phase this is, however, not the only solution and not the global minimum. The
nontrivial solution is found by means of $\partial V_{\text{1L}}/\partial
v^2=0$. Formulas \eqref{1loop_potential} and \eqref{Delta} give
\begin{multline}
0=\frac{\partial V_{\text{1L}}}{\partial v^2}=\frac12(M^2+\delta
M^2-\mu^2)+\frac12(\lambda+\delta\lambda)v^2\\
+\frac i2\int\frac{d^4k}{(2\pi)^4}\left[
2\lambda\Delta_{11}^{\phi\he\phi}(k)-
\frac{\frac{\partial}{\partial
v^2}\det\Delta_{22}^{-1}(k)}{\det\Delta_{22}^{-1}(k)}\right].
\label{aux}
\end{multline}

The momentum integral in Eq. \eqref{aux} is only formal and implicitly
involves a~convenient regularization. In deriving Eq.
\eqref{aux} we already assumed that the regularization allows to
change the sign of the integration variable so that the contribution of the
particles circulating in the loop is the same as that of antiparticles. For the
later calculation of the scalar field propagator it will also be useful to
be able to shift the integration variable. In concrete computations we shall
use the analytic integration over energy (frequency) in combination with
dimensional regularization around $d=3$ space dimensions
\cite{Andersen:2005yk}.

Eq. \eqref{aux} may be rewritten in a~particularly useful way.
With a~decent use of Eq. \eqref{propagator_tree_level} we find
(for the sake of brevity, we temporarily drop out the arguments of the
propagators as well as the integration measure)
\begin{multline*}
\frac{\partial}{\partial v^2}\det\Delta_{22}^{-1}=
\frac{\partial}{\partial
v^2}\left(\Delta_{22}^{-1\phi\he\phi}\Delta_{22}^{-1\he\phi\phi}-
\Delta_{22}^{-1\phi\phi}\Delta_{22}^{-1\he\phi\he\phi}\right)\\
=-2\lambda\left(
\Delta_{22}^{-1\phi\he\phi}+\Delta_{22}^{-1\he\phi\phi}-\Delta_{22}^{-1\phi\phi}\right)
\end{multline*}
so that
\begin{multline*}
-\int\frac{\frac{\partial}{\partial
v^2}\det\Delta_{22}^{-1}}{\det\Delta_{22}^{-1}}=
2\lambda\int\left(\Delta_{22}^{\he\phi\phi}+\Delta_{22}^{\phi\he\phi}
+\Delta_{22}^{\phi\phi}\right)\\
=2\lambda\int\left(2\Delta_{22}^{\phi\he\phi}+\Delta_{22}^{\phi\phi}\right).
\end{multline*}
Eq. \eqref{aux} thus acquires the form
\begin{multline}
\mu^2-(M^2+\delta
M^2)-(\lambda+\delta\lambda)v^2\\
=2i\lambda\int\frac{d^4k}{(2\pi)^4}\left[
\Delta_{11}^{\phi\he\phi}(k)+2\Delta_{22}^{\phi\he\phi}(k)+\Delta_{22}^{\phi\phi}(k)
\right]
\label{gap_equation}
\end{multline}
Even though this result has been derived by a~not very transparent
manipulation, it could have been expected: It is equivalent, up to an overall
factor, to the requirement that the one-particle-irreducible tadpole
contributions to the vacuum expectation value of $\phi_2$ vanish. For the
diagrammatic representation of the individual terms see Appendix
\ref{Sec:diagrammar}.

For the concrete calculations, we adopted the following renormalization
conditions. The scalar mass was renormalized by subtracting the whole one-loop
correction at $\mu=0$ and $v=0$ [see also Eq. \eqref{propagator_normal_phase}
later], i.e.,
\begin{equation}
\delta M^2=-3\lambda\Lambda^{3-d}\int\frac{d^d\vec k}{(2\pi)^d}
\frac1{\sqrt{\vec k^2+M^2}},
\label{mass_counterterm}
\end{equation}
where the integral is regularized by minimal subtraction in $d$ dimensions and
$\Lambda$ is the renormalization scale. This procedure guarantees that the
parameter $M$ keeps its interpretation as the physical (pole) mass of the
scalar at zero chemical potential.

The coupling constant $\lambda$ was renormalized by modified minimal
subtraction ($\overline{\text{MS}}$) so that
$$
\delta\lambda=\frac{3\lambda^2}{4\pi^2}\left(\frac2\epsilon-\gamma+\log4\pi\right),
$$
where $\epsilon=3-d$ is the expansion parameter of dimensional regularization
\footnote{This is of course a standard textbook result --- see, for instance,
Eq. (11.76) in Ref. \cite{Peskin:1995ev}. It should be noted, however, that we
use a slightly different regularization than usual, namely the dimensional
regularization in \emph{three} spatial dimensions.}. (The pole does not depend
on the chemical potential and hence could be extracted from the analytically
calculable loop integral at $\mu=0$.) The renormalization scale $\Lambda$ was
set equal to the characteristic scale of the system: we chose $\Lambda=M$ in
the normal phase and $\Lambda=\mu$ in the Bose--Einstein condensed phase. Since
the phase transition occurs at $\mu=M$ (see the following discussion), the
subtraction point is a~connected function of the chemical potential.

For sake of numerical solution the gap equation \eqref{gap_equation} was
rewritten, with the help of Eq. \eqref{Delta} and upon the integration over
frequencies, as
\begin{multline}
v^2=\frac{\mu^2-M^2}\lambda-\int\frac{d^3\vec k}{(2\pi)^3}\left\{
\frac1{\sqrt{\vec k^2+M^2+\lambda v^2}}\right.\\
+\left.\frac2{E_+(\vec k)+E_-(\vec k)}\left[1+
\frac{\vec k^2+M^2+\frac32\lambda v^2-\mu^2}{E_+(\vec k)E_-(\vec
k)}\right]\right\}\\
-\frac{\delta M^2+\delta\lambda\,v^2}{\lambda}.
\label{v_numerical}
\end{multline}
This equation was solved iteratively with the initial ansatz given by the
tree-level value, $v_0=\sqrt{\frac{\mu^2-M^2}{\lambda}}$. To implement the
$\overline{\text{MS}}$ scheme within the numerical computation, we subtracted
the pole part in the form
$$
\delta\lambda=-\frac{3\lambda^2}{4\pi^2}\log\frac{\Lambda^2}{M^2}+
3\lambda^2\Lambda^{3-d}\int\frac{d^d\vec k}{(2\pi)^d}\frac1{(\vec
k^2+M^2)^{3/2}}.
$$
The integral here reproduces the pole and is, in fact, equal to the full one-loop
correction to $\lambda$ at $\mu=0$ and $v=0$. Upon the subtraction of $\delta
M^2$ and $\delta\lambda$, the integral in \eqref{v_numerical} is rendered
finite and was evaluated approximately with a~simple cutoff.

\begin{figure}
\begin{center}
\scalebox{1}{\includegraphics{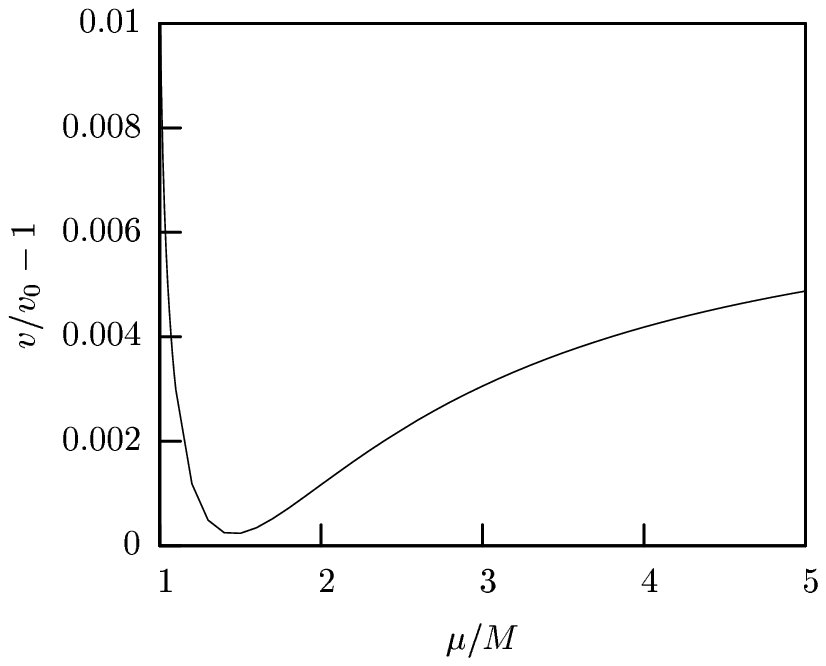}}
\end{center}
\caption{Relative increase of the vacuum expectation value $v$ with respect to
the tree-level value $v_0=\sqrt{\frac{\mu^2-M^2}{\lambda}}$. The numerical data
were obtained with the renormalized coupling $\lambda$ set to $1$.}
\label{Fig:vev}
\end{figure}

The numerical results are displayed in Fig. \ref{Fig:vev}. We emphasize the
fact that \emph{the phase transition to the Bose--Einstein condensed phase
happens at the chemical potential equal to the renormalized scalar mass}. This
is of course not surprising from the physical point of view and in fact is an
important check of consistency of our calculations.

This conclusion is also easily proved analytically. Just set $v=0$ and then
immediately observe that the right hand side of Eq. \eqref{gap_equation} is
exactly canceled by the counterterm $\delta M^2$ i.e., $v=0$ is a~solution of
Eq. \eqref{gap_equation} exactly for $\mu=M$.

Finally, note that the convergence of our iterative solution of the gap
equation guarantees that the found solution is, as required, a~(at least local)
minimum of the effective potential.

\section{Scalar propagator}
\label{Sec:scalar_propagator} The one-loop-corrected propagator $\D$ is
obtained as a~second functional derivative of the one-loop effective action
\eqref{1loop_action},
$$
\D^{-1}_{ij}(x,y)=\frac{\delta^2\Gamma_{\text{1L}}}{\delta\he\Phi_i(x)\delta\Phi_j(y)}.
$$
After differentiating the logarithm of the determinant we arrive at
\begin{multline*}
\D^{-1}_{ij}(x,y)=\Delta^{-1}_{ij}(x,y)\\
+\frac
i2\Tr\left[\Delta\frac{\delta^2\Delta^{-1}}{\delta\he\Phi_i(x)\delta\Phi_j(y)}-
\Delta\frac{\delta\Delta^{-1}}{\delta\he\Phi_i(x)}\Delta\frac{\delta\Delta^{-1}}
{\delta\Phi_j(y)}\right]
\end{multline*}
plus counterterms. Here `$\Tr$' denotes the full functional trace, over both
spacetime and internal degrees of freedom.

The derivatives of the inverse bare propagator $\Delta^{-1}$ are nothing but
the cubic and the quartic interaction vertex. They are local and may be
represented by the coupling matrices $T^{(i)}$ and $Q^{(ij)}$, defined as
\begin{align*}
\frac{\delta\Delta^{-1}_{kl}(s,t)}{\delta\he\Phi_i(x)}&=T^{(i)}_{kl}\delta^4(s-t)\delta^4(s-x)\\
\frac{\delta^2\Delta^{-1}_{kl}(s,t)}{\delta\he\Phi_i(x)\delta\Phi_j(y)}&=
Q^{(ij)}_{kl}\delta^4(s-t)\delta^4(s-x)\delta^4(t-y).
\end{align*}
The explicit formulas for the matrices $T^{(i)}$ and $Q^{(ij)}$ are listed in the
Appendix \ref{Sec:coupling_matrices}. With this notation, the inverse propagator
acquires the form
\begin{multline}
\D^{-1}_{ij}(p)=\Delta^{-1}_{ij}(p)\\
+\frac i2\int\frac{d^4k}{(2\pi)^4}\tr
\left[\Delta(k)Q^{(ij)}-\Delta(k)T^{(i)}\Delta(p+k)T^{(j)\dagger}\right]
\label{full_propagator}
\end{multline}
plus counterterms. The symbol `$\tr$' here refers to a~trace over internal
degrees of freedom only, that is, over the flavor and Nambu spaces.

\subsection{Normal phase}
When the chemical potential is small enough, the system is in the normal phase:
$v=0$ and the symmetry is not spontaneously broken. In this phase the one-loop
correction to the propagator is particularly simple. The bare propagator
$\Delta$ is diagonal in both the flavor and the Nambu space, the only nonzero
components being
\begin{align*}
\Delta_{ij}^{\phi\he\phi}(p)&=[(p_0+\mu)^2-\vec p^2-M^2]^{-1},\\
\Delta_{ij}^{\he\phi\phi}(p)&=[(p_0-\mu)^2-\vec p^2-M^2]^{-1}.
\end{align*}

Since in the normal phase there is no cubic coupling $T^{(i)}$,
the final formula for the one-loop propagator is easily found to be
\begin{multline}
\D^{-1}_{ij\phi\he\phi}(p)=\bigl[(p_0+\mu)^2-\vec p^2-(M^2+\delta M^2)\bigr]\delta_{ij}\\
-6i\lambda\delta_{ij}\int\frac{d^4k}{(2\pi)^4}\frac1{(k_0+\mu)^2-\vec k^2-M^2}.
\label{propagator_normal_phase}
\end{multline}
The other nonzero component of the propagator, $\D_{\he\phi\phi}$, is related
to Eq. \eqref{propagator_normal_phase} by the first of the identities,
\begin{equation}
\begin{split}
\D_{\he\phi\phi}(p)&=\D_{\phi\he\phi}(-p),\\
\D_{\he\phi\he\phi}(p)&=\D^*_{\phi\phi}(p)=\D^*_{\phi\phi}(-p).
\end{split}
\label{propagator_relations}
\end{equation}
which follow from the very
definition of the propagator.

Note that the loop correction in Eq. \eqref{propagator_normal_phase} is
momentum-independent, as usual in a~$\phi^4$ theory. The renormalization is
therefore trivial and corresponds to a~mere redefinition of the mass.

Since in dimensional regularization we may freely shift the integration
variable, the loop integral in Eq. \eqref{propagator_normal_phase} is
$\mu$-independent and thus is completely canceled by the mass counterterm
\eqref{mass_counterterm}. Therefore, in the normal phase, the scalar field
propagator gets no correction and the mass spectrum simply consists of two
doubly degenerate levels at $M\pm\mu$ \cite{Miransky:2001tw,Schaefer:2001bq}.

\subsection{Bose--Einstein condensed phase}
When the scalar condenses i.e., $v>0$, we have to work with the full formulas
\eqref{Delta}. The matrix propagator $\D$ has altogether 16 components (four flavor
times four Nambu). Fortunately, several of them are actually zero due to the
clever choice of the vacuum, Eq. \eqref{vacuum_choice}. With this choice, the
unbroken flavor $\mathrm{U(1)}_Q$ symmetry is generated by the matrix
$Q=\tfrac12(\openone+\tau_3)$. This means that the upper flavor component of
$\phi$, $\phi_1$, carries the unbroken charge while the lower component $\phi_2$
does not. The charge conservation then immediately implies that the propagator
is diagonal in the flavor space, i.e., $\D_{12}(p)=\D_{21}(p)=0$. Moreover, the
propagator of $\phi_1$ is diagonal in the Nambu space, that is,
$\D_{11}^{\phi\phi}(p)=\D_{11}^{\he\phi\he\phi}(p)=0$. [All these relations
could, of course, also be demonstrated explicitly by a~proper analysis of the
formula \eqref{full_propagator}.]

The remaining nonzero components of the propagator are strongly constrained by
Eq. \eqref{propagator_relations} so that only three of them are independent:
$\D_{11}^{\phi\he\phi}$, $\D_{22}^{\phi\he\phi}$ and $\D_{22}^{\phi\phi}$.

\subsubsection{Propagator of $\phi_1$}
\label{Sec:propPhi1} The only independent component of $\D_{11}$ is
$\D_{11}^{\phi\he\phi}$. A~straightforward application of Eq.
\eqref{full_propagator} yields
\begin{widetext}
\begin{multline}
\D_{11}^{-1\phi\he\phi}(p)=(p_0+\mu)^2-\vec p^2-(M^2+\delta
M^2)-(\lambda+\delta\lambda)v^2\\-2i\lambda\int\frac{d^4k}{(2\pi)^4}
\left[2\Delta_{11}^{\phi\he\phi}(k)+\Delta_{22}^{\phi\he\phi}(k)\right]-2i\lambda^2v^2
\int\frac{d^4k}{(2\pi)^4}\Delta_{11}^{\phi\he\phi}(p+k)\left[\Delta_{22}^{\phi\he\phi}(k)+
\Delta_{22}^{\phi\phi}(k)+\Delta_{22}^{\he\phi\he\phi}(k)+\Delta_{22}^{\he\phi\phi}(k)\right].
\label{prop_phi1}
\end{multline}
\end{widetext}
The first integral represents the tadpoles and is momentum-independent.
A~qualitative difference in comparison with the normal phase comes in the second
integral, which contains the contribution of the cubic vertices. The
diagrammatic representation of all terms in the propagator is given in Appendix
\ref{Sec:diagrammar}.

The propagator $\D_{11}^{\phi\he\phi}(p)$ is expected to have a~massless pole
corresponding to the Goldstone boson. Indeed, a~short glance at Eq.
\eqref{Delta} reveals that at tree level, when $\lambda v^2=\mu^2-M^2$, there
is a~gapless pole in $\Delta_{11}^{\phi\he\phi}(p)$. The field $\phi_1$
annihilates a~particle with dispersion relation $E=\sqrt{\vec p^2+\mu^2}-\mu$,
i.e., a~type-II Goldstone boson. The other excitation, annihilated by
$\he\phi_1$, with a~tree-level gap $2\mu$ is manifested as a~pole in
$\D_{11}^{\he\phi\phi}(p)$.

The analytic proof of the existence of a~massless pole in
$\D_{11}^{\phi\he\phi}(p)$ is provided in Appendix \ref{Sec:poleD11}. Here we
just observe that once the relation $\D_{11}^{-1\phi\he\phi}(0)=0$ is proved, we
may subtract it from $\D_{11}^{-1\phi\he\phi}(p)$ to obtain a~convenient
expression
\begin{multline}
\D_{11}^{-1\phi\he\phi}(p)=p_0^2+2\mu p_0-\vec
p^2\\
-2i\lambda^2v^2\int\frac{d^4k}{(2\pi)^4}\left[\Delta_{11}^{\phi\he\phi}(p+k)-
\Delta_{11}^{\phi\he\phi}(k)\right]\\
\times\left[\Delta_{22}^{\phi\he\phi}(k)+
\Delta_{22}^{\phi\phi}(k)+\Delta_{22}^{\he\phi\he\phi}(k)
+\Delta_{22}^{\he\phi\phi}(k)\right].
\label{prop_phi1_renormalized}
\end{multline}
Note that the propagator no longer depends explicitly on the renormalization
counterterms. Also, the momentum-independent tadpole graphs disappeared and the
remaining integral is finite and perfectly well defined even when the
regularization is removed.

In order to establish the radiative corrections to the type-II Goldstone boson
dispersion relation, we performed the integration over frequency and expanded
the result in powers of the external momentum $p$, up to order $p_0$ and $\vec
p^2$, respectively. The final formula is rather cumbersome, so we do not write
it out and instead report just the result of the numerical integration of the
remaining momentum integral.

\begin{figure}
\begin{center}
\scalebox{1}{\includegraphics{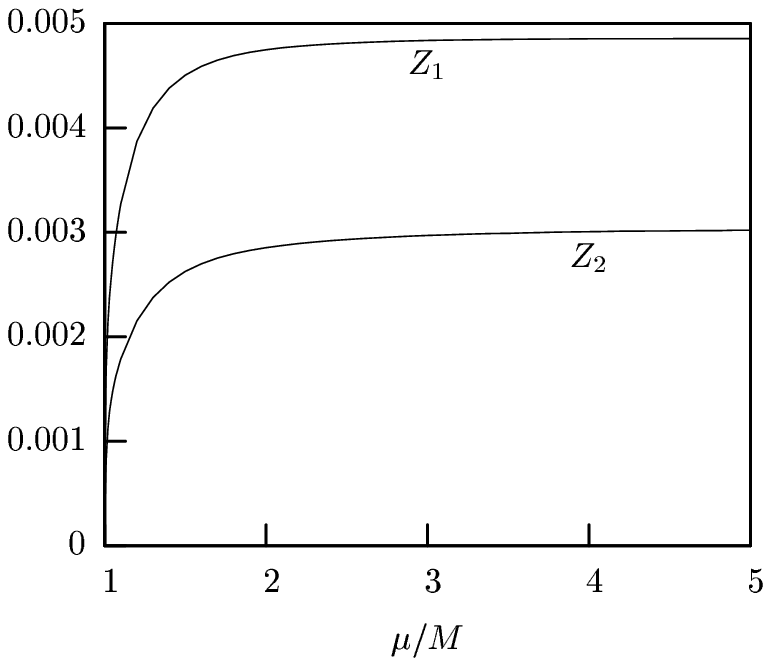}}
\end{center}
\caption{Renormalization constants of the type-II Goldstone boson propagator.
The numerical results indicate that their dependence on the chemical potential
is linear near the phase transition and saturates at large $\mu$. The data were
obtained with $\lambda=1$.}
\label{Fig:disp_rel_GB1}
\end{figure}

At the leading order of the momentum power expansion, the inverse propagator
reads
$$
\D_{11}^{-1\phi\he\phi}(p)=2\mu p_0(1+Z_1)-\vec p^2(1+Z_2).
$$
The coefficients $Z_1,Z_2$ determine the renormalization of both the wave
function and the dispersion relation of the Goldstone boson. Their dependence
on the chemical potential is displayed in Fig. \ref{Fig:disp_rel_GB1}.

Now we turn our attention to the massive partner of the type-II Goldstone
boson. Analogously to Ref. \cite{Schaefer:2001bq}, we shall calculate just the
leading order of its dispersion relation, i.e., its gap. The corresponding
propagator, $\D_{11}^{\he\phi\phi}$, is obtained from Eq.
\eqref{prop_phi1_renormalized} via the relations \eqref{propagator_relations}.
Next we set $\vec p=\vec 0$. However, the explicit form of the tree-level
propagators, Eq. \eqref{Delta}, reveals that
$\Delta_{11}^{\phi\he\phi}(k_0-2\mu,\vec k)=\Delta_{11}^{\he\phi\phi}(k_0,\vec
k)$ so that the difference of propagators in the loop correction to
$\D_{11}^{\phi\he\phi}(-p_0,\vec 0)$ exactly vanishes at $p_0=2\mu$, see Eq.
\eqref{prop_phi1_renormalized}. As a~result, the renormalized gap is equal to
its tree-level value $2\mu$ and \emph{receives no radiative corrections}.

\subsubsection{Propagator of $\phi_2$}
\label{Sec:propPhi2}
In this case, Eq. \eqref{full_propagator} yields, after a~reasonable
application of the rules \eqref{propagator_relations},
\begin{widetext}
\begin{multline}
\D_{22}^{-1\phi\he\phi}(p)=(p_0+\mu)^2-\vec p^2-(M^2+\delta
M^2)-2(\lambda+\delta\lambda)v^2-2i\lambda\int\frac{d^4k}{(2\pi)^4}
\left[\Delta_{11}^{\phi\he\phi}(k)+2\Delta_{22}^{\phi\he\phi}(k)\right]\\
-2i\lambda^2v^2\int\frac{d^4k}{(2\pi)^4}\Delta_{11}^{\phi\he\phi}(k)
\Delta_{11}^{\phi\he\phi}(p+k)\\
-8i\lambda^2v^2\int\frac{d^4k}{(2\pi)^4}\left\{\Delta_{22}^{\phi\he\phi}(p+k)
\left[\Delta_{22}^{\phi\he\phi}(k)+\Delta_{22}^{\phi\phi}(k)+
\Delta_{22}^{\he\phi\he\phi}(k)+\frac12\Delta_{22}^{\he\phi\phi}(k)\right]
+\Delta_{22}^{\phi\phi}(p+k)\Delta_{22}^{\he\phi\he\phi}(k)\right\},
\label{prop_phi2hephi2}
\end{multline}
\begin{multline}
\D_{22}^{-1\phi\phi}(p)=-(\lambda+\delta\lambda)v^2-
2i\lambda\int\frac{d^4k}{(2\pi)^4}\Delta_{22}^{\phi\phi}(k)
-2i\lambda^2v^2\int\frac{d^4k}{(2\pi)^4}
\Delta_{11}^{\phi\he\phi}(k)\Delta_{11}^{\phi\he\phi}(p+k)\\
-8i\lambda^2v^2\int\frac{d^4k}{(2\pi)^4}\left\{
\Delta_{22}^{\phi\phi}(p+k)\left[
\Delta_{22}^{\phi\he\phi}(k)+\frac12\Delta_{22}^{\phi\phi}(k)+
\Delta_{22}^{\he\phi\he\phi}(k)+\Delta_{22}^{\he\phi\phi}(k)\right]
+\Delta_{22}^{\phi\he\phi}(p+k)\Delta_{22}^{\phi\he\phi}(k)\right\}.
\label{prop_phi2phi2}
\end{multline}
\end{widetext}

Here the propagator is not diagonal in the Nambu space and the situation is
thus more complicated. We cannot look for the zeros of the individual
components of the inverse propagator. Instead, we have to solve the condition
$\det\D_{22}^{-1}(p)=0$, which seems a~rather formidable task in view of Eqs.
\eqref{prop_phi2hephi2} and \eqref{prop_phi2phi2}.

Fortunately, we need not calculate the determinant explicitly, at least to
prove the existence of a~Goldstone boson. Recalling the relations
\eqref{propagator_relations}, the determinant at zero momentum simplifies to
$$
\det\D_{22}^{-1}(0)=\left[\D_{22}^{-1\phi\he\phi}(0)\right]^2
-\left|\D_{22}^{-1\phi\phi}(0)\right|^2.
$$
The secular equation thus reduces to
$$
\D_{22}^{-1\phi\he\phi}(0)=\pm\left|\D_{22}^{-1\phi\phi}(0)\right|.
$$
In view of Eq. \eqref{propagator_tree_level} we see that the existence of
a~Goldstone boson is proved once we verify the relation
\begin{equation}
\D_{22}^{-1\phi\he\phi}(0)=\D_{22}^{-1\phi\phi}(0).
\label{pole_phi2}
\end{equation}
A~detailed proof of this statement is provided in Appendix \ref{Sec:poleD22}.

The analysis of Eqs. \eqref{prop_phi2hephi2} and \eqref{prop_phi2phi2} would
yield the radiative correction to the type-I Goldstone boson dispersion
relation, i.e., the speed of sound in the Bose--Einstein condensate. We do not
perform the detailed calculation here since our main concern was the type-II
Goldstone boson studied above. Moreover, such a~straightforward approach is
unnecessarily complicated because the one-loop correction to the speed of sound
has been determined by Andersen using a~different method
\cite{Andersen:2005yk}. In the following, we shall instead concentrate on
another specific property of the superfluid phonon in our system --- its decay.

\section{Phonon decay}
\label{Sec:phonon_decay} Upon inspection of Eq.
\eqref{disp_rel_tree_level_exact}, the full tree-level dispersion relation of
the type-I Goldstone boson (or, the phonon) turns out to be convex. This fact
has far-reaching consequences for it kinematically allows the phonon to decay.
(A~similar, convex, dispersion law is found e.g. for the superfluid phonon in
the Bose--Einstein condensed weakly interacting Bose gas
\cite{Andersen:2003qj}.) This in turn strongly affects the transport properties
of the system such as the shear viscosity \cite{Manuel:2004iv}.

Here we calculate the two-particle decay rate of the phonon at the leading
order of perturbation theory. Such a~quantity could in principle be extracted,
via the optical theorem, from the imaginary part of the one-loop propagator
calculated in Sec. \ref{Sec:propPhi2}. This would, however, be rather
complicated for two reasons. First, the phonon is, in the propagator of
$\phi_2$, entangled with the massive `radial' mode. Second, we would have to
deal with the full momentum dependence of the propagator; the low-momentum
limit studied before would not be sufficient for our purpose. Since the
upcoming calculation of the phonon decay rate involves a~few rather nonstandard
pieces, we regard it instructive to perform it step by step, using ordinary
perturbation theory.

Before starting the calculation we briefly comment why we have chosen to study
such a~particular process. First, as has already been mentioned above, the
low-energy properties of the Goldstone bosons are essential for the
thermodynamics of the system at low temperature. Two factors are determinant
for the possibility of the decay of the Goldstone bosons: The conservation of
the unbroken $\mathrm{U(1)}_Q$ charge and phase space (kinematic) restrictions.
The type-II Goldstone boson, $G$, carries the unbroken charge. Since at low
energy it can only decay to gapless particles, the only processes allowed by
charge conservation are $G\to G\pi$, $G\to G\pi\pi$, etc. (Here and in the
following, $\pi$ denotes the type-I Goldstone boson i.e., the phonon.) Taking
into account the dispersion relations \eqref{disp_rel_tree_level}, these decay
modes, however, fail to satisfy the energy and momentum conservation
simultaneously. Thus, the type-II Goldstone boson is stable at low momentum.

On the other hand, the phonon may decay, by the processes $\pi\to\pi\pi$,
$\pi\to\pi\pi\pi$ etc. Only the two processes explicitly written here receive
nonzero contribution at the first order of perturbation theory. As will become
clear later, the three-particle decay is suppressed, at least in certain ranges
of momentum and chemical potential. We shall therefore calculate just the decay
rate for the process $\pi\to\pi\pi$ and comment on the relevance of our result
afterwards.

\subsection{Vacuum transition amplitudes}
The decay rate cannot be straightforwardly deduced from the Lagrangian
\eqref{Lagrangian} because it is expressed in terms of the scalar field $\phi$
rather than the physical modes $\pi, G$ and their massive counterparts. The
problem is that the kinetic term in the Lagrangian cannot be diagonalized at
the level of fields. Indeed, Eq. \eqref{Delta} shows that the diagonalizing
unitary transformation is necessarily energy-dependent. We therefore have to
calculate the three-point Green's function of $\phi$ and use the LSZ reduction
formula to extract the physical scattering amplitude.

Alternatively, we may directly calculate the $S$-matrix; the couplings of
$\phi$ to the physical states follow from the K\"all\'en--Lehmann
representation of the propagator. Let $A(x)$ and $B(x)$ be two local bosonic
operators. The general form of the K\"all\'en--Lehmann representation of the
time-ordered Green's function, $\D^{AB}(x,y)=-i\bra0T\{A(x)B(y)\}\ket0$,
reads, in the momentum representation,
\begin{multline}
\D^{AB}(\vec p,\omega)=(2\pi)^3\sum_n\left[ \frac{\bra0A(0)\ket{n,\vec
p}\bra{n,\vec p}B(0)\ket0}{\omega-E(\vec
p)+i\epsilon}\right.\\
\left.-\frac{\bra0B(0)\ket{n,-\vec p}\bra{n,-\vec p}A(0)\ket0}{\omega+E(\vec
p)-i\epsilon}\right], \label{Kallen}
\end{multline}
where $E(\vec p)$ is the energy of the Hamiltonian eigenstate $\ket{n,\vec p}$.
The summation index $n$ is discrete for one-particle intermediate states, and
continuous for multiparticle states. Also, the states are assumed to be
normalized as $\bra{n,\vec p}m,\vec k\rangle=\delta_{mn}\delta^3(\vec p-\vec
k)$. Note that we use the general K\"all\'en--Lehmann representation
\eqref{Kallen} rather than the covariant form usual in literature
\cite{Peskin:1995ev}.

The poles in the propagator of $\phi_2$ correspond to the dispersion relations
given in Eq. \eqref{disp_rel_tree_level_exact}. The contribution of the
massless pole at $\omega=E_-(\vec p)-i\epsilon$ is
\begin{widetext}
\begin{multline*}
\Delta_{22}^{\text{pole}}(\vec p,\omega)=\frac1{2E_-(\vec p)}\frac1{E_-^2(\vec
p)-E_+^2(\vec p)}
\frac1{\omega-E_-(\vec p)+i\epsilon}\\
\times
\left(\begin{array}{cc}
[E_-(\vec p)-\mu]^2-\vec p^2-M^2-2\lambda v^2 & \lambda v^2\\
\lambda v^2 & [E_-(\vec p)+\mu]^2-\vec p^2-M^2-2\lambda v^2
\end{array}\right).
\end{multline*}
This expression immediately yields the transition amplitudes
\begin{equation}
\begin{split}
\bra0\phi_2(0)\ket{\pi(\vec p)}&=\frac1{(2\pi)^{3/2}}\frac1{\sqrt{2E_-(\vec p)}}
\frac1{\sqrt{E_+^2(\vec p)-E_-^2(\vec p)}}
\Bigl\{-[E_-(\vec p)-\mu]^2+\vec p^2+M^2+2\lambda v^2\Bigr\}^{1/2},\\
\bra0\he\phi_2(0)\ket{\pi(\vec p)}&=-\frac1{(2\pi)^{3/2}}\frac1{\sqrt{2E_-(\vec p)}}
\frac1{\sqrt{E_+^2(\vec p)-E_-^2(\vec p)}}
\Bigl\{-[E_-(\vec p)+\mu]^2+\vec p^2+M^2+2\lambda v^2\Bigr\}^{1/2}.
\end{split}
\label{transition_amplitudes}
\end{equation}
\end{widetext}
The relative sign of the amplitudes follows from the nondiagonal part of the
propagator. Note that the amplitudes are particularly simple in the limit
$v\to0$ (i.e., at the phase boundary) where $E_{\pm}(\vec p)=\sqrt{\vec
p^2+M^2}\pm\mu=\epsilon_{\vec p}\pm\mu$. It follows that, in this limit,
\begin{equation}
\bra0\phi_2(0)\ket{\pi(\vec p)}=\frac1{(2\pi)^{3/2}\sqrt{2\epsilon_{\vec
p}}},\quad
\bra0\he\phi_2(0)\ket{\pi(\vec p)}=0.
\label{trans_ampl_limit}
\end{equation}
Thus the result typical for Lorentz-invariant theories is recovered in the
normal phase.

At the first order of perturbation theory, the amplitude for the two-particle
decay of $\pi$ follows from the Lagrangian \eqref{Lagrangian} upon shifting
$\phi_2$ by its vacuum expectation value, $v/\sqrt2$. The relevant cubic terms
read
$$
-\sqrt2\lambda v(\he\phi_2\phi_2\he\phi_2+\he\phi_2\phi_2\phi_2).
$$
These two terms generate three diagrams each, contributing to the given decay
process, that differ just by a~permutation of the lines. The result for the
$S$-matrix element is
$$
S=-(2\pi)^4\delta^4(p-k-q)i\mathcal M\frac1{\left[(2\pi)^{3/2}\right]^3},
$$
where the scattering amplitude $\mathcal M$ is given by
\begin{multline}
-i\mathcal M\frac1{\left[(2\pi)^{3/2}\right]^3}=-2\sqrt2i\lambda v\\
\times\Bigl[
\bra{\pi(\vec k)}\he\phi_2\ket0\bra{\pi(\vec
q)}\he\phi_2\ket0\bra0\phi_2\ket{\pi(\vec p)}\Bigr.\\
+\bra{\pi(\vec k)}\he\phi_2\ket0\bra{\pi(\vec
q)}\phi_2\ket0\bra0\he\phi_2\ket{\pi(\vec p)}\\
+\bra{\pi(\vec k)}\phi_2\ket0\bra{\pi(\vec
q)}\he\phi_2\ket0\bra0\he\phi_2\ket{\pi(\vec p)}\\
+\bra{\pi(\vec k)}\phi_2\ket0\bra{\pi(\vec
q)}\phi_2\ket0\bra0\he\phi_2\ket{\pi(\vec p)}\\
+\bra{\pi(\vec k)}\phi_2\ket0\bra{\pi(\vec
q)}\he\phi_2\ket0\bra0\phi_2\ket{\pi(\vec p)}\\
\Bigl.+\bra{\pi(\vec k)}\he\phi_2\ket0\bra{\pi(\vec
q)}\phi_2\ket0\bra0\phi_2\ket{\pi(\vec p)}\Bigr].
\label{scattering_amplitude}
\end{multline}
For legibility we omitted the arguments `$(0)$' of the field operators. Also,
the initial state is labeled by the momentum $\vec p$, the momenta of the
products are $\vec k,\vec q$.


\subsection{Kinematics}
Our analysis of the two-particle decay is further complicated by the fact that
the Lorentz invariance is broken by the chemical potential. We therefore have
to evaluate the decay rate as a~function of the momentum of the decaying
particle. After the common trick of squaring the $S$-matrix in finite volume
and performing the trivial three-momentum integration we arrive at the formula
for the decay rate,
$$
\Gamma=\frac12\frac1{4\pi^2}\int\vec k^2d^3\vec k\,|\mathcal M|^2
\delta\bigl(E_-(\vec p)-E_-(\vec k)-E_-(\vec q)\bigr).
$$
The factor $\tfrac12$ arises from the fact that the two particles in the final
state are identical. The $\delta$-function fixes the moduli of the momenta
$\vec k$ and $\vec q$ as a~function of the scattering angle $\theta$, which
measures the deflection of $\vec k$ from the direction of $\vec p$. The $|\vec
q|$ is given by $|\vec q|=\sqrt{\vec p^2+\vec k^2-2|\vec p||\vec k|\cos\theta}$
and $|\vec k|$ is then fixed by energy conservation. Due to the form of the
dispersion, Eq. \eqref{disp_rel_tree_level_exact}, the solution of the energy
conservation condition is, however, rather involved, so we work it out
analytically only in a~special case (see Sec. \ref{Sec:decay_rate}). In the
general case, the decay rate is computed numerically.

\begin{figure}
\begin{center}
\scalebox{1}{\includegraphics{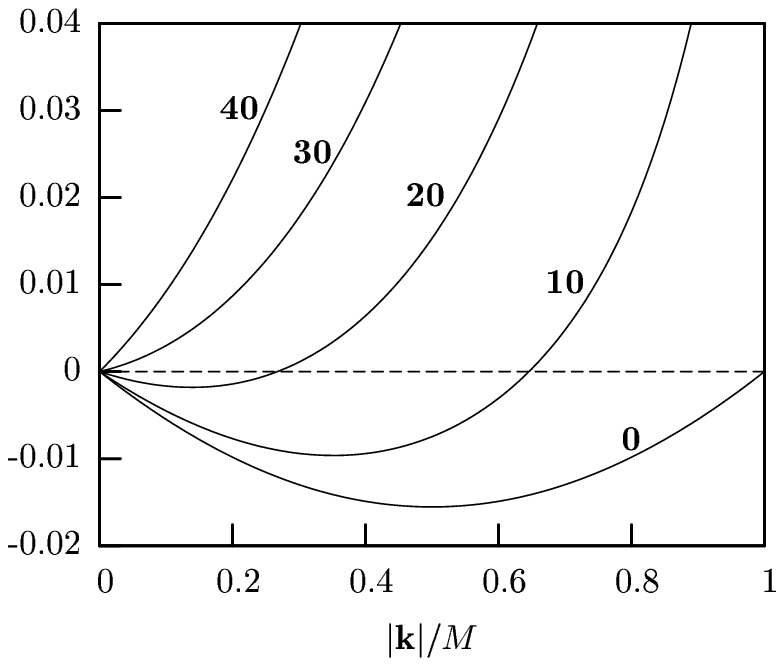}}
\end{center}
\caption{Sample plot of the expression $-E_-(\vec p)+E_-(\vec k)+E_-(\vec q)$
(in units of $M$) as a~function of $|\vec k|$, for $\mu=2M$ and $|\vec p|=M$.
The curves are labeled by the value of the angle $\theta$ in degrees.
The critical angle for the chosen set of parameters is $\theta_m\approx26.3\deg$.}
\label{Fig:energy_conservation}
\end{figure}

Using the chain rule to differentiate $E_-(\vec q)$ in the $\delta$-function,
we arrive at the final formula for the decay rate,
\begin{multline}
\Gamma=\frac1{4\pi}\int_0^{\theta_m}\sin\theta\,d\theta\,\vec k^2(\theta)\\
\times\frac{|\mathcal M|^2}
{\left|E'_-\bigl(\vec k(\theta)\bigr)+E'_-\bigl(\vec q(\theta)\bigr)
\frac{|\vec k(\theta)|-|\vec p|\cos\theta}
{|\vec q(\theta)|}\right|},
\label{decay_rate}
\end{multline}
where $\theta_m$ is the maximum angle to which $\pi$ can decay at a~given
energy. Fig. \ref{Fig:energy_conservation} displays the expression $-E_-(\vec
p)+E_-(\vec k)+E_-(\vec q)$ as a~function of $|\vec k|$, for several values of
the angle $\theta$. The critical angle $\theta_m$ is apparently determined by
the condition that the derivative of the argument of the energy-conserving
$\delta$-function [i.e., the expression in the denominator in Eq.
\eqref{decay_rate}] vanishes at $\vec k=\vec 0$. Since for $\vec k=\vec 0$ both
energy and momentum conservation require $\vec q=\vec p$, this reduces to
$\cos\theta_m=E'_-(0)/E'_-(\vec p)$. Using explicitly the dispersion relation
\eqref{disp_rel_tree_level_exact}, we have
\begin{multline}
\cos\theta_m=\frac{E_-(\vec p)}{|\vec p|}
\sqrt{\frac{\mu^2-M^2}{3\mu^2-M^2}}\\
\times\left[1-\frac{2\mu^2}{\sqrt{4\mu^2\vec p^2+(3\mu^2-M^2)^2}}\right]^{-1}.
\label{critical_angle}
\end{multline}

Formula \eqref{critical_angle} admits several particularly simple limits. For
$|\vec p|\to0$ we find $\theta_m=0$. This is easily understandable: At low
momentum the phonon dispersion relation is linear and the decay thus
kinematically forbidden. For $|\vec p|\to\infty$, on the other hand,
we have $\theta_m=\arccos\sqrt{\frac{\mu^2-M^2}{3\mu^2-M^2}}$. Finally, for
$\mu\to M+$ at fixed momentum i.e., at the phase transition, $\theta_m=\pi/2$.
This shows that one has to be careful when performing these limits. Near the
phase transition there may be two small scales, $|\vec p|$ and $\mu-M$, and the
result depends on their hierarchy.

\subsection{Decay rate}
\label{Sec:decay_rate} The phonon decay rate may be determined analytically in
the limit $v\to0$. Strictly speaking, the following calculation is justified
only for $\vec p^2\gg\lambda v^2$. However, later the analytic result will be
checked by a direct numerical integration of the full formula
\eqref{decay_rate}. For the time being, we simply set $v=0$ everywhere except,
of course, the overall factor $v$ in the amplitude $\mathcal M$.

Eq. \eqref{trans_ampl_limit} shows that only the first term survives in Eq.
\eqref{scattering_amplitude} so that
$$
\mathcal M=2\sqrt2\lambda v\frac1{\sqrt{2\epsilon_{\vec p}}}
\frac1{\sqrt{2\epsilon_{\vec k}}}\frac1{\sqrt{2\epsilon_{\vec q}}}.
$$
Setting $\mu=M$ in Eq. \eqref{disp_rel_tree_level_exact} gives $E_{\pm}(\vec
p)=\epsilon_{\vec p}\pm M$, hence $E_-'(\vec p)=|\vec p|/\epsilon_{\vec p}$ and
Eq. \eqref{decay_rate} reduces to
\begin{multline}
\Gamma=\frac{\lambda^2v^2}{4\pi\epsilon_{\vec
p}}\int_0^{\pi/2}\sin\theta\,d\theta\\
\times\frac{\vec k^2(\theta)}{|\vec k(\theta)|(\epsilon_{\vec k(\theta)}+
\epsilon_{\vec q(\theta)})-|\vec p|\epsilon_{\vec k(\theta)}\cos\theta}.
\label{mid_stage_decay}
\end{multline}

With the simplified dispersion relation, the energy and momentum conservation
may easily be solved to yield
$$
|\vec k|=\frac{2M|\vec p|E_+(\vec p)\cos\theta}{E_+^2(\vec p)-\vec
p^2\cos^2\theta},\quad
E_-(\vec k)=\frac{2M\vec p^2\cos^2\theta}{E_+^2(\vec p)-\vec
p^2\cos^2\theta}.
$$
The angular integration in Eq. \eqref{mid_stage_decay} is then elementary and
we arrive at the final formula for the decay rate,
\begin{equation}
\Gamma=\frac{\lambda^2v^2}{4\pi}\frac{|\vec p|}{\epsilon_{\vec
p}(\epsilon_{\vec p}+M)}.
\label{decay_rate_limit}
\end{equation}

\begin{figure}
\begin{center}
\scalebox{1}{\includegraphics{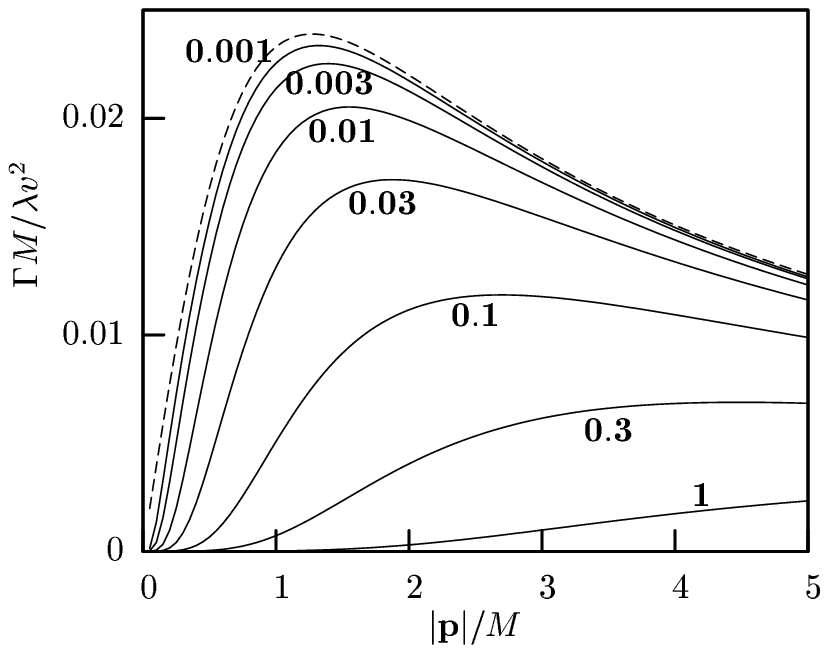}}
\end{center}
\caption{Decay rate for the process $\pi\to\pi\pi$ as a~function of momentum
(solid lines). The numerical data were obtained with $\lambda=1$. The curves
are labeled by the dimensionless parameter $\mu/M-1$. The dashed line
corresponds to the analytic result, Eq. \eqref{decay_rate_limit}. The decay
rates for different values of the chemical potential are scaled by the factor
$\lambda v^2$ so that their convergence to the analytic result is comparable.}
\label{Fig:decay_rate}
\end{figure}

Such a~dependence of the decay rate on the momentum of the initial phonon might
have been expected: The decay rate tends to zero at both small and large
momenta, where the dispersion relation becomes linear. A~comparison with the
exact numerical integration of Eq. \eqref{decay_rate} is displayed in Fig.
\ref{Fig:decay_rate}.

\section{Conclusions}
In this paper we studied spontaneous symmetry breaking by Bose--Einstein
condensation in scalar $\phi^4$ theory with two charged scalar flavors. Our
primary goal was to refine our previous results \cite{Brauner:2005di} by
including radiative corrections.

Within the one-loop calculation we verified that the Bose--Einstein
condensation sets on at the chemical potential equal to the renormalized scalar
mass (at $\mu=0$). This assertion, though perfectly physical, may seem
surprising within the approximation used since, as is well known
\cite{Peskin:1995ev}, in massless $\phi^4$ theory the one-loop correction draws
the scalar field expectation value off the origin to a~nonzero value. The
spontaneous symmetry breaking thus achieved, however, eventually turns out to
be an artifact of the one-loop effective potential and disappears upon the
renormalization group improvement.

Fortunately, such a~fake vacuum apparently does not occur in our case for the
phase transition point is exactly where it should be. This gives us hope that
the renormalization group improvement, however important for critical
phenomena, should not change the conclusions of this paper qualitatively.

In Sec. \ref{Sec:propPhi1}, we determined the radiative corrections to the
dispersion relations of the type-II Goldstone boson and its massive
counterpart. The numerical results suggest that the dependence of the
renormalization constants $Z_1$ and $Z_2$ on the chemical potential saturates
at large $\mu$. Since in this region the two massive modes in the model become
very heavy, the asymptotic behavior of the type-II Goldstone boson dispersion
relation should be governed by a~low-energy effective theory for the Goldstone
bosons, presumably analogous to that for the nonrelativistic ferromagnet
\cite{Leutwyler:1994gf,Hofmann:1998pp}.

Furthermore, we showed that the massive counterpart of the type-II Goldstone
boson has gap $2\mu$, which receives no radiative corrections. This assertion
has a~simple physical interpretation. Recall that $\mu$ is, originally,
a~chemical potential associated with the $\mathrm{U(1)}$ symmetry of the
Lagrangian \eqref{Lagrangian}. Once this symmetry is spontaneously broken,
$\mu$ may be reinterpreted as a~chemical potential of the unbroken
$\mathrm{U(1)}_Q$. Since $\phi_1$ and $\he\phi_1$ carry the charges $Q=\pm1$,
the corresponding excitations differ by $2\mu$ in energy.

In a~sense, this is a~remnant of the Goldstone theorem: It guarantees
nonrenormalization of the Goldstone boson mass to all orders in the loop
expansion. Once the chemical potential is switched on, the excitation
annihilated by $\he\phi_1$ can no longer be massless, but the Goldstone theorem
still assures that its gap does not renormalize.

In Sec. \ref{Sec:phonon_decay}, we determined the two-particle decay rate of
the superfluid phonon. Such a~quantity as well as the very possibility of the
phonon decay may be important for the transport phenomena in the
Color-Flavor-Locked quark matter with a~kaon condensate \cite{Manuel:2004iv}.
As shown in Refs. \cite{Bedaque:2001je,Kaplan:2001qk,Warringa:2006dk}, the CFL
phase with a~meson condensate appears to be energetically preferred to the
simple CFL phase, thus making our results of a~possible relevance for the
phenomenology of compact stars.

To summarize, in this paper we investigated spontaneous symmetry breaking
within the linear sigma model at the one-loop level. We verified the saturation
of the Nielsen--Chadha inequality \eqref{Nielsen_Chadha} and hence also the
counting rule proposed (and proved at the tree level) in our previous work
\cite{Brauner:2005di}. We thus fulfilled one part of the program outlined in
Ref. \cite{Brauner:2005di}. The other, more ambitious part is to extend the
results achieved so far to other relativistic systems with finite chemical
potential. Our recent work shows that at least the type-II Goldstone bosons may
be described, to a large extent, in a model-independent way. These results will
be reported elsewhere.

\appendix
\section{Cubic and quartic coupling matrices}
\label{Sec:coupling_matrices}
In this appendix the explicit form of the coupling matrices $T^{(i)}$ and
$Q^{(ij)}$ is given. The indices $\phi$ and $\he\phi$ refer to the upper and
lower components of the Nambu doublet, respectively.
\begin{align*}
T^{(i)}_{kl\phi}&=\left(
\begin{array}{cc}
-2\lambda(\delta_{il}\phi_k+\delta_{kl}\phi_i) & 0\\
-2\lambda(\delta_{il}\he\phi_k+\delta_{ik}\he\phi_l) &
-2\lambda(\delta_{ik}\phi_l+\delta_{kl}\phi_i)
\end{array}\right),\\
T^{(i)}_{kl\he\phi}&=\left(
\begin{array}{cc}
-2\lambda(\delta_{ik}\he\phi_l+\delta_{kl}\he\phi_i) &
-2\lambda(\delta_{il}\phi_k+\delta_{ik}\phi_l)\\
0 & -2\lambda(\delta_{il}\he\phi_k+\delta_{kl}\he\phi_i)
\end{array}\right),\\
Q^{(ij)}_{kl\phi\he\phi}&=\left(
\begin{array}{cc}
-2\lambda(\delta_{jk}\delta_{il}+\delta_{kl}\delta_{ij}) & 0\\
0 & -2\lambda(\delta_{ik}\delta_{jl}+\delta_{kl}\delta_{ij})
\end{array}\right),\\
Q^{(ij)}_{kl\phi\phi}&=\left(
\begin{array}{cc}
0 & 0\\
-2\lambda(\delta_{ik}\delta_{jl}+\delta_{jk}\delta_{il}) & 0
\end{array}\right),\\
Q^{(ij)}_{kl\he\phi\he\phi}&=\left(
\begin{array}{cc}
0 & -2\lambda(\delta_{ik}\delta_{jl}+\delta_{jk}\delta_{il})\\
0 & 0
\end{array}\right),\\
Q^{(ij)}_{kl\he\phi\phi}&=\left(
\begin{array}{cc}
-2\lambda(\delta_{ik}\delta_{jl}+\delta_{kl}\delta_{ij}) & 0\\
0 & -2\lambda(\delta_{jk}\delta_{il}+\delta_{kl}\delta_{ij})
\end{array}\right).
\end{align*}

\section{Diagrammatic representation}
\label{Sec:diagrammar} In the following, we summarize the diagrammatic
representation of the individual terms in the effective potential as well as
the propagators. The flavor components $\phi_1$ and $\phi_2$ are denoted by
oriented dashed and dotted lines, respectively. Since the $\D_{22}$ propagator
is not diagonal in the Nambu space, it has to carry two arrows, specifying the
operators it connects at its ends. The bare propagators \eqref{Delta} are
represented by the following lines.
\begin{gather*}
\Delta_{11}^{\phi\he\phi}=
\parbox{15\unitlength}{%
\includegraphics{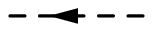}}\qquad\qquad
\Delta_{11}^{\he\phi\phi}=
\parbox{15\unitlength}{%
\rotatebox{180}{\includegraphics{fg34}}}\\
\Delta_{22}^{\phi\he\phi}=
\parbox{15\unitlength}{%
\includegraphics{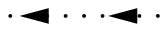}}\qquad\qquad
\Delta_{22}^{\phi\phi}=
\parbox{15\unitlength}{%
\includegraphics{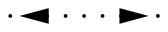}}\\
\Delta_{22}^{\he\phi\he\phi}=
\parbox{15\unitlength}{%
\includegraphics{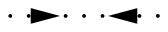}}\qquad\qquad
\Delta_{22}^{\he\phi\phi}=
\parbox{15\unitlength}{%
\includegraphics{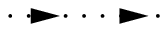}}\\
\end{gather*}

We start with the interaction vertices. These all come from the
$-\lambda(\he\phi\phi)^2$ term in the Lagrangian. Upon shifting $\phi_2$ by
$v/\sqrt2$, the interaction yields the following vertices,
\begin{gather*}
\parbox{15\unitlength}{%
\includegraphics{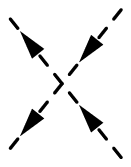}}=-4i\lambda,\quad
\parbox{15\unitlength}{%
\includegraphics{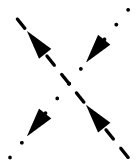}}=-2i\lambda,\\
\parbox{15\unitlength}{%
\includegraphics{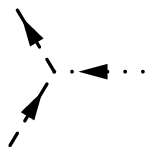}}\,=
\parbox{15\unitlength}{%
\includegraphics{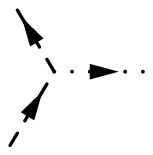}}\,=-\sqrt2i\lambda v,\\
\parbox{15\unitlength}{%
\includegraphics{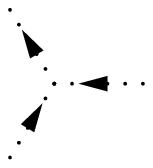}}\,=
\parbox{15\unitlength}{%
\includegraphics{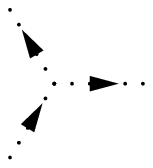}}\,=-2\sqrt2i\lambda v,\quad
\parbox{15\unitlength}{%
\includegraphics{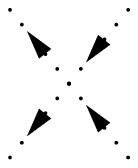}}=-4i\lambda.
\end{gather*}

One-particle irreducible contributions to the
vacuum expectation value of the scalar field $\phi_2$, see Eq.
\eqref{gap_equation}:
\begin{multline*}
\parbox{12\unitlength}{%
\includegraphics{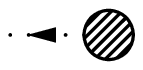}}\,=
\parbox{12\unitlength}{%
\includegraphics{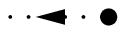}}\,+
\parbox{17\unitlength}{%
\includegraphics{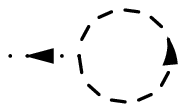}}\\+
\parbox{17\unitlength}{%
\includegraphics{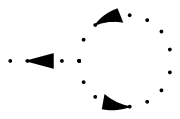}}\,+\frac12\,
\parbox{17\unitlength}{%
\includegraphics{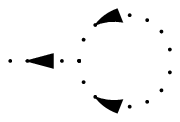}}
\end{multline*}
We have explicitly indicated the symmetry factor $\tfrac12$ at the last graph.
Note that in our formalism all such factors are automatically generated by the
functional differentiation of the effective action.

Next is the diagrammatic representation of the propagators. The loop graphs
are in one-to-one correspondence with the terms in the formulas \eqref{prop_phi1},
\eqref{prop_phi2hephi2} and \eqref{prop_phi2phi2}.
\begin{widetext}
\begin{equation*}
\D_{11}^{-1\phi\he\phi}=
\Delta_{11}^{-1\phi\he\phi}+
\parbox{15\unitlength}{%
\includegraphics{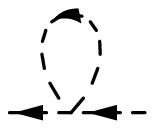}}\,+
\parbox{15\unitlength}{%
\includegraphics{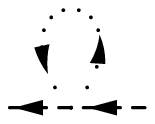}}\,+
\parbox{20\unitlength}{%
\includegraphics{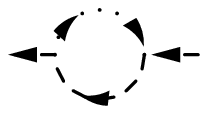}}\,+
\parbox{20\unitlength}{%
\includegraphics{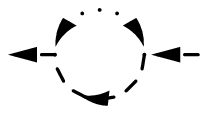}}\,+
\parbox{20\unitlength}{%
\includegraphics{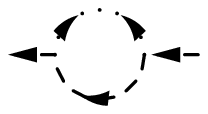}}\,+
\parbox{20\unitlength}{%
\includegraphics{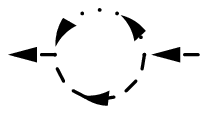}}
\end{equation*}
\begin{multline*}
\D_{22}^{-1\phi\he\phi}=
\Delta_{22}^{-1\phi\he\phi}+
\parbox{15\unitlength}{%
\includegraphics{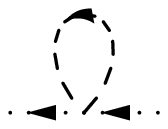}}\,+
\parbox{15\unitlength}{%
\includegraphics{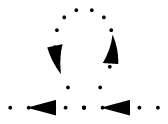}}\,+
\parbox{20\unitlength}{%
\includegraphics{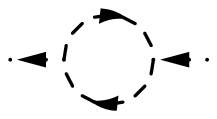}}\\+
\parbox{20\unitlength}{%
\includegraphics{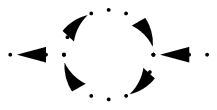}}\,+
\parbox{20\unitlength}{%
\includegraphics{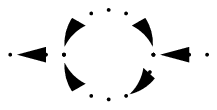}}\,+
\parbox{20\unitlength}{%
\includegraphics{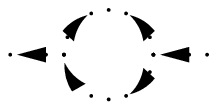}}\,+
\frac12\,\parbox{20\unitlength}{%
\includegraphics{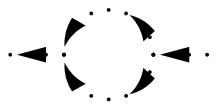}}\,+
\parbox{20\unitlength}{%
\includegraphics{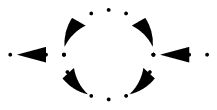}}
\end{multline*}
\begin{multline*}
\D_{22}^{-1\phi\phi}=
\Delta_{22}^{-1\phi\phi}+
\frac12\,\parbox{15\unitlength}{%
\includegraphics{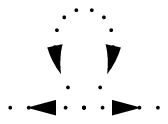}}\,+
\parbox{20\unitlength}{%
\includegraphics{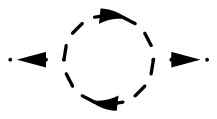}}\\+
\parbox{20\unitlength}{%
\includegraphics{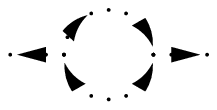}}\,+
\frac12\,\parbox{20\unitlength}{%
\includegraphics{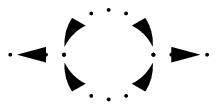}}\,+
\parbox{20\unitlength}{%
\includegraphics{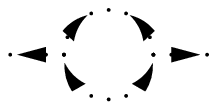}}\,+
\parbox{20\unitlength}{%
\includegraphics{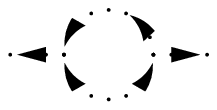}}\,+
\parbox{20\unitlength}{%
\includegraphics{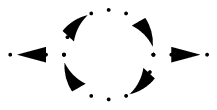}}
\end{multline*}

\section{Proof of the existence of gapless poles}
\label{Sec:poles}
Here we provide a~detail proof of the existence of massless poles in the
propagators $\D_{11}$ and $\D_{22}$. The key ingredients are the formulas
\eqref{prop_phi1}, \eqref{prop_phi2hephi2} and \eqref{prop_phi2phi2}, as well
as the gap equation \eqref{gap_equation}.

\subsection{Pole in $\D_{11}$}
\label{Sec:poleD11}
We are going to prove that $\D_{11}^{-1\phi\he\phi}(0)=0$. Eq,
\eqref{prop_phi1} together with the expressions \eqref{Delta} for the bare
propagators give
\begin{multline}
\D_{22}^{-1\phi\he\phi}(0)=\mu^2-(M^2+\delta M^2)-(\lambda+\delta\lambda)v^2
-4i\lambda\int\frac{d^4k}{(2\pi)^4}\frac1{(k_0+\mu)^2-\epsilon_{\vec k}^2-\lambda
v^2}
-2i\lambda\int\frac{d^4k}{(2\pi)^4}\frac{(k_0-\mu)^2-\epsilon_{\vec k}^2-2\lambda v^2}
{[k_0^2-E_+^2(\vec k)][k_0^2-E_-^2(\vec k)]}\\
-2i\lambda^2v^2\int\frac{d^4k}{(2\pi)^4}\frac1{(k_0+\mu)^2-\epsilon_{\vec k}^2-\lambda v^2}
\frac{[(k_0+\mu)^2-\epsilon_{\vec k}^2-\lambda v^2]+[(k_0-\mu)^2-\epsilon_{\vec k}^2-\lambda v^2]}
{[k_0^2-E_+^2(\vec k)][k_0^2-E_-^2(\vec k)]},
\label{first_step}
\end{multline}
where we abbreviate $\epsilon_{\vec k}=\sqrt{\vec k^2+M^2}$. Next we expand the
gap equation \eqref{gap_equation}, using the explicit form of the propagators
\eqref{Delta},
\begin{multline*}
\mu^2-(M^2+\delta M^2)-(\lambda+\delta\lambda)v^2=
2i\lambda\int\frac{d^4k}{(2\pi)^4}\frac1{(k_0+\mu)^2-\epsilon_{\vec
k}^2-\lambda v^2}\\
+4i\lambda\int\frac{d^4k}{(2\pi)^4}\frac{(k_0+\mu)^2-\epsilon_{\vec k}^2-2\lambda v^2}
{[k_0^2-E_+^2(\vec k)][k_0^2-E_-^2(\vec k)]}+
2i\lambda^2v^2\int\frac{d^4k}{(2\pi)^4}\frac1{[k_0^2-E_+^2(\vec k)][k_0^2-E_-^2(\vec
k)]}.
\end{multline*}
This expression is used to substitute for
$\mu^2-(M^2+\delta M^2)-(\lambda+\delta\lambda)v^2$ in Eq. \eqref{first_step}, which finally
yields
$$
\D_{11}^{-1\phi\he\phi}(0)=-2i\lambda\int\frac{d^4k}{(2\pi)^4}\left\{
\frac1{(k_0+\mu)^2-\epsilon_{\vec k}^2-\lambda v^2}-
\frac{(k_0-\mu)^2-\epsilon_{\vec k}^2-2\lambda v^2}
{[k_0^2-E_+^2(\vec k)][k_0^2-E_-^2(\vec k)]}\left[1-\frac{\lambda v^2}
{(k_0+\mu)^2-\epsilon_{\vec k}^2-2\lambda v^2}\right]\right\}.
$$
Using the identity
$$
\det\Delta_{22}^{-1}(k)=[k_0^2-E_+^2(\vec k)][k_0^2-E_-^2(\vec k)]=
[(k_0+\mu)^2-\epsilon_{\vec k}^2-2\lambda v^2][(k_0-\mu)^2-\epsilon_{\vec k}^2-2\lambda
v^2]-(\lambda v^2)^2,
$$
which follows from Eq. \eqref{Delta}, it is now
straightforward to show that the integrand in the last expression vanishes.

\subsection{Pole in $\D_{22}$}
\label{Sec:poleD22}
In this case we are going to prove Eq. \eqref{pole_phi2}. We shall use a~more
symbolic method, without referring to the explicit form of the bare propagators
\eqref{Delta}. Upon a~comparison of Eqs. \eqref{prop_phi2hephi2} and
\eqref{prop_phi2phi2} we can see that several terms immediately cancel out.
After a~mild rearrangement we are to prove the identity
\begin{equation}
\mu^2-(M^2+\delta M^2)-(\lambda+\delta\lambda)v^2=2i\lambda\int
\left(\Delta_{11}^{\phi\he\phi}+2\Delta_{22}^{\phi\he\phi}-\Delta_{22}^{\phi\phi}\right)
+4i\lambda^2v^2\int\underbrace{\left(\Delta_{22}^{\phi\he\phi}\Delta_{22}^{\he\phi\phi}-
\Delta_{22}^{\phi\phi}\Delta_{22}^{\phi\phi}\right)}_{\det\Delta_{22}}.
\label{second_step}
\end{equation}
For the sake of legibility we dropped out the integration measure and arguments
of the propagators. Also, we used of the fact that $\Delta_{22}^{\phi\phi}=
\Delta_{22}^{\he\phi\he\phi}$.

On the right hand side of Eq. \eqref{second_step}, we rewrite
$\lambda v^2\det\Delta_{22}$ as $\lambda
v^2/\det\Delta_{22}^{-1}=\Delta_{22}^{\phi\phi}$ and immediately arrive at an
expression identical to the right hand side of Eq. \eqref{gap_equation}.
Eq. \eqref{second_step} is thus proved and hence also the
existence of a~massless pole in the propagator $\D_{22}$.
\end{widetext}

\begin{acknowledgments}
The author is indebted to J. Ho\v{s}ek and J. Novotn\'{y} for helpful
discussions and/or critical reading of the manuscript. The present work was
supported in part by the Institutional Research Plan AV0Z10480505, and by the
GACR grants No. 202/06/0734 and 202/05/H003.
\end{acknowledgments}

\end{document}